# Estudos de cenários de implantação de um imposto ambiental no transporte aéreo no Brasil

Carolina Barbosa Resende
Alessandro V. M. Oliveira⟶
Instituto Tecnológico de Aeronáutica, São José dos Campos, Brasil
⟶ Autor correspondente. Instituto Tecnológico de Aeronáutica. Praça Marechal Eduardo Gomes, 50. 12.280-250 - São José dos Campos, SP - Brasil.
E-mail: alessandro@ita.br.

**Resumo**: Nos últimos anos, o tema aquecimento global e emissão de gases do efeito estufa tem estado cada vez mais na mídia. Esse tema tem levantado diversas bandeiras, mostrando preocupação com o futuro, e buscando alternativas para uma vida mais sustentável ao longo dos anos. Ao estudar emissão de gases do efeito estufa, um dos principais emissores de tais gases é a queima de combustíveis, na qual veículos de transporte que dependem de gás e óleo se enquadram. Neste quesito, o transporte aéreo por meio de aeronaves é uma das fontes emissoras de tais gases. Visando reduzir a emissão de gases, uma das formas de se fazer isso é através da redução do consumo de combustível, para isso aeronaves teriam que ser mais eficientes, ou oferta e demanda de voos seria reduzida. No entanto, e se o consumo de combustível pelas aeronaves fosse taxado? Quais seriam os efeitos disso diante do transporte aéreo? Seria essa uma das alternativas para reduzir as emissões? Para entender essa relação da taxação e uma possível redução no consumo de combustível, um estudo foi desenvolvido pelo Instituto Tecnológico de Aeronáutica (ITA), usando um modelo econométrico para entender como a demanda seria afetada pela mudança nas tarifas aéreas. Além disso, uma simulação de possíveis impostos ambientais sob os valores das passagens aéreas foi realizada para se analisar a resposta da demanda e se ter uma ideia se essa taxação realmente resolveria os problemas das emissões.

**Palavras-chave**: transporte aéreo, companhias aéreas, econometria.

## I. Introdução

A indústria do transporte aéreo é uma das fontes emissoras de gases do efeito estufa. Isso porque aeronaves dependem de combustível a base de petróleo para taxiarem e voarem, e para isso, há a queima de combustível, o que consequentemente leva à emissão de gases do efeito estufa. Segundo Schumann (2002), mais de 70% das emissões das aeronaves são compostas por dióxido de carbono ($CO_2$), o qual é um dos principais gases causadores do efeito estufa. Não apenas isso, mas há estudos que sugerem que a aviação é responsável por mais de 3% de todas as emissões globais de $CO_2$ (Simões e Schaeffer, 2005).

No ano de 2015, ocorreu a 21a conferência das partes da convenção-quadro das nações unidas (COP21). Essa conferência abordou as emissões de gases como um de seus principais temas, e acabou gerando o acordo de Paris. Esse acordo visava combater as mudanças climáticas, e limitar o aumento da temperatura global, como parte deste acordo, foi previsto a redução das emissões de gases do efeito estufa. Na COP24, que aconteceu em 2018, essas mudanças do acordo de Paris ainda estavam sendo discutidas, mostrando a importância do tema.

Como uma forma de reduzir as emissões de gases do efeito estufa, alguns países da União Europeia (UE) estudaram a aplicação de um imposto ambiental sobre o consumo de combustível, tal imposto foi chamado por eles de taxação ambiental (EUROSTAT, 2013). No entanto, tal taxação pode se tornar um custo extra para as empresas aéreas, e essas empresas podem arcar com esses custos extras, repassar parte desses custos para os passageiros, ou repassar todos esses custos para o passageiro (Resende e Oliveira, 2017). Ao repassar os custos para o passageiro, ocorre um aumento no preço das passagens aéreas, e tal aumento pode fazer com que passageiros desistam da viagem, impactando diretamente a demanda pelos voos.

Ao gerar uma redução na demanda devido ao aumento das tarifas, a empresa pode optar ou não por reduzir as frequências de voos ou tomar outras atitudes, o que poderia ou não levar a uma redução do consumo de combustível, e consequentemente das emissões. No entanto, o impacto na demanda pode afetar além da empresa aéreas, mas pode afetar a economia do país de uma forma geral devido ao impacto no turismo e vendas de determinados locais (Resende e Oliveira, 2017).

Tendo essas informações em mente, uma pergunta foi levantada pelo grupo de pesquisas do ITA. Aplicar uma taxação ambiental sobre o consumo de combustível de voos domésticos das empresas aéreas brasileiras vai reduzir as emissões de gases do efeito estufa? Para responder a essas perguntas, a pesquisa realizada precisou compreender a resposta da demanda a um aumento do preço das passagens aéreas, o que foi possível devido a uma análise econométrica do modelo estudado, e consequente simulação da taxação ambiental sob o consumo de combustível. A resposta final a esta pergunta é que se torna difícil dizer se a taxação resolve ou não os problemas das emissões, pois é necessário saber se a empresa aérea realmente iria agir modificando a malha e frota para redução do consumo de combustível, e também pois não se sabe a reação real da demanda. Iria esse passageiro desistir da viagem e realmente reduzir as emissões de gases? Ou iria esse passageiro mudar o modo de transporte, e ao invés de ir de avião iria de ônibus ou carro, apenas transferindo as emissões de um modo de transporte para outro? Este capítulo irá mostrar como se chegou a essas conclusões, e qual seria o impacto da taxação na demanda pelo transporte aéreo.

## II. Revisão da Literatura

Dentre a literatura internacional, Hofer, Dresner e Windle (2010) mostraram que para o mercado norte americano, uma redução na demanda pelo transporte aéreo não reduzia as emissões de gases necessariamente, na verdade poderia inclusive gerar um aumento nas emissões. Isso porque segundo os autores, poderia haver uma migração para outro modal de transporte, como o rodoviário, e que o aumento das emissões pelo novo modal rodoviário poderia superar a economia do modal aéreo. Ou seja, neste caso a taxação poderia gerar um efeito contrário ao esperado.

No entanto, mesmo a taxação podendo ter efeitos contrários, diversos países vêm estudando a prática. Por exemplo, Andersen (2010) e Convery, Dunne e Joyce (2014), mostrou que a Irlanda vem estudando a aplicação de uma taxa entre 15 e 30 euros por tonelada de $CO_2$ emitido. Para chegar à estimativa de tonelada de $CO_2$ emitido, a ICAO criou uma calculadora online capaz de estimar essas tal valor com base em distância e outras estimativas para determinados voos. Segundo Iannone (2011), essas taxas geraram um aumento de até 3,5% nas tarifas aéreas finais.

Outro tema abordado ao se tratar de taxações ambientais ao consumo de combustível é saber como essa taxação seria aplicada, quem receberia, dentre outros. Por exemplo, segundo Iannone (2011), a UE tinha a intenção de taxar qualquer voo que sobrevoasse seu território, independente de ter origem ou destino na UE. Porém essa taxação pode ser devida apenas aos locais de origem e destino, ou podem ser devidas a outros locais e órgão a depender do acordo fechado.

Ainda com relação a taxação ambiental, Brouwer, Brander e Van Beukering (2008) e Jou e Chen (2015) disseram que passageiros podem se sensibilizar com a causa da redução das emissões, o que os tornam dispostos a pagar mais pelas passagens aéreas. Isso mostra que pode não haver uma reação linear entre o aumento dos preços devido a taxação e a redução da demanda. Além disso, essa reação pode varia conforme o tipo de passageiro, rota voada, dentre outros.

Além de se buscar entender a taxação ambiental, também é necessário entender como o preço do combustível pode ser impactado. Fatores como a prática do hedging, a mudança na malha e na frota da empresa aérea, a formação de alianças estratégicas, dentre outros, podem ser motivos de variação nos custos com combustível para a empresa aérea.

Uma outra alternativa não abordada anteriormente é o chamado fuel tankering, ou tanqueamento (Morrel e Swan, 2006; e Ryerson e Hansen, 2013) e o controle de velocidade e altitude do voo. Tanqueamento é o ato de abastecer as aeronaves em locais com custos mais baixos, seja devido a menores impostos, subsídios, ou outros motivos (Fregnani, Müller e Correia, 2013). O tanqueamento deve ser feito com cautela, uma vez que ao se abastecer a aeronave em determinados locais devido aos custos, a aeronave irá decolar com peso superior ao necessário para a viagem devido ao peso do combustível em excesso, o que faz com que o consumo de combustível para a viagem seja superior ao necessário. Esse fato deve ser bem analisado antes de aplicado, uma vez que o combustível gasto a mais na viagem aumenta as emissões, e esse gasto a mais pode superar a economia gerada pelo tanqueamento. Segundo Fregnani et al. (2013), o tanqueamento é limitado ao peso máximo de decolagem e a capacidade de tanque da aeronave.

O controle da velocidade e altitude do voo por sua vez é uma alternativa interessante para a redução de emissões. Babikian, Lukachko e Waitz (2002) discutem o aumento da eficiência no consumo de combustível relacionada ao controle de velocidade e altitude do voo..

## III. Estudo Nacional

Utilizando o caso hipotético de cenário de taxação ambiental no Brasil, Resende e Oliveira (2017) simularam a mudança da demanda com relação aos possíveis valores das passagens aéreas após a implantação de um imposto. Um modelo econométrico foi utilizado para dar base à construção de cenários que quantificassem tal impacto, analisando a elasticidade-preço da demanda, ou em outras palavras, a variação da demanda devido a alterações nos preços. Esse estudo utilizou dados entre janeiro de 2003 e dezembro de 2013. Período este em que houve instabilidade econômica entre outubro de 2006 e julho de 2007, também chamado de "apagão aéreo" (Oliveira, Ferreira e Salgado, 2011), e no último trimestre de 2008 devido a bolha do setor imobiliário dos EUA (Freitas, 2009).

O estudo de Resende e Oliveira (2017) traz duas principais hipóteses com relação ao impacto da taxação ambienta sobre a demanda do transporte aéreo. A primeira hipótese é de que a reação da demanda não é constante ou linear à taxação ambiental, e de que quanto maior a taxação ambiental, maior a redução da demanda, de forma não linear. A segunda hipótese é de que o perfil dos passageiros ou da rota impactam na reação da demanda, ou seja, passageiros com diferentes motivos de viagens, e rotas com diferentes perfis de passageiros possuem impacto na demanda diferentes entre si.

Tendo as hipóteses em mente, diversos cenários foram testados e comparados pelo modelo econométrico, visando a escolha do modelo com variáveis que melhor explicava os resultados. Sendo assim, a Figura 1 apresenta o modelo escolhido, no qual (+) gera uma relação direta positiva com relação à demanda aérea, por exemplo se a variável analisada aumenta ou é existente, a demanda também aumenta; (-) gera uma relação direta contrária em relação à demanda, por exemplo se a variável aumenta ou é existente, a demanda diminui; e (NS) significa que a variável não é significante para a análise da variação da demanda.

**Tabela 1 - Resumo dos resultados**

| Variável | Demanda |
|---|---|
| Densidade populacional | + |
| Renda da população | + |
| Tarifa aérea | - |
| Existência de empresa *codeshare* | - |
| Período do "apagão aéreo" | - |
| Período de crise econômica | NS |
| Presença de nova empresa *low cost* | + |
| Interação tarifa aérea e proporção de passageiros que já fizeram a mesma roda em modal diferente | - |
| Interação tarifa aérea e proporção de passageiros voando a negócios | + |
| Interação tarifa aérea e existência de nova empresa *low cost* na rota | - |

Fonte: Resende e Oliveira (2017)

Com base nos resultados obtidos pelos autores, temos que o período de crise econômica global, relacionado à bolha do setor imobiliário nos EUA, foi a única variável no qual não se notou diferença significativa na demanda. Um dos motivos que explica essa não significância é por ser um período pequeno frente a base de dados coletada, e a existência de outras crises financeiras que afetaram o setor aéreo brasileiro na base de dados.

No entanto, os resultados mostram que um aumento na densidade populacional gera um aumento na demanda. O que faz sentido, uma vez que em cidades mais povoadas, existem mais pessoas dispostas a viajar, por isso o efeito positivo. Da mesma maneira a variável renda impacta a demanda de forma positiva, ou seja, quanto maior a renda da população, maior a probabilidade das pessoas viajarem e utilizarem o transporte aéreo para tal.

A existência de empresas codeshare representa empresas que cooperam entre si. Isto é, o passageiro compra a passagem com a empresa A, porém voa no avião da empresa B. Isso possibilita com que o passageiro faça conexões mais eficientes, e evita compra de passagens avulsas, o que pode gerar um preço mais elevado para a passagem final. No caso estudado, codeshare é a presença das empresas TAM e Varig entre os anos de 2003 e 2005. Essa variável apresentou resultados negativos no modelo, ou seja, a demanda era reduzida nesse período para rotas em que se operavam voos dessas empresas.

O período chamado "apagão aéreo" foi outra variável utilizada no modelo com efeito negativo, ou seja, nesse período a demanda era reduzida no transporte aéreo. O que está em acordo com Oliveira et al. (2011). Além disso, a existência de nova empresa low cost no período é considerado como a presença da empresa Azul linhas aéreas, a qual entrou no mercado brasileiro em 2008. Essa é a única nova low cost no período, e sua presença possui um efeito positivo, dizendo que a chegada da Azul gerou um aumento na demanda pelo transporte aéreo brasileiro.

De maneira oposta, a tarifa aérea possui relação negativa com demanda, o que significa que com um aumento nas tarifas aéreas, espera-se uma redução nas demandas. Essa análise faz sentido ao se pensar na curva da demanda, onde maiores preços fazem com que menos pessoas busquem comprar o produto, e menores preços faz com que mais pessoas busquem comprar o produto. Porém, para melhor compreensão dessa análise das tarifas, a tarifa aérea foi interagida com outras variáveis. A primeira interação foi realizada com a proporção de passageiros que já fizeram a mesma rota utilizando outro meio de transporte, essa interação gerou um resultado negativo. Esse resultado mostra que no caso da possibilidade dos passageiros em fazer a mesma rota utilizando outro meio de transporte, tal como carro ou ônibus, a demanda reduz conforme a tarifa aumenta. Isso mostra que os passageiros podem estar dispostos a manter a viagem e mudar o meio de transporte utilizado, ou podem desistir da viagem devido ao aumento do preço das passagens aéreas.

A segunda interação se deu entre as tarifas aéreas e a proporção de passageiros viajando a negócios. Essa interação teve efeito positivo, indicando que as tarifas aéreas não impactam a demanda de negócios da mesma maneira que impactam passageiros a turismo. Em outras palavras, a existência de passageiros a negócio pode reduzir a elasticidade da tarifa, ou seja, passageiros a negócios são menos afetados a um aumento do preço das passagens.

A terceira e última interação se deu entre tarifas aéreas e presença de nova empresa low cost. Essa interação teve efeito negativo e representa que no caso da existência da nova empresa low cost, um aumento nas tarifas aéreas gera uma queda na demanda. Porém essa reação de queda da demanda é mais forte nos casos onde não há empresas low cost. Isso mostra que a nova low cost possibilita aos passageiros manter a viagem e trocar a empresa aérea com a qual voa, de maneira a pagar valores aceitáveis e dentro de seu orçamento, se comparado a rotas onde não há low cost.

Sabendo como se dá a elasticidade-preço da demanda, foi possível simular um aumento nas tarifas aéreas decorrente da taxação ambiental, para então avaliar a resposta da demanda. Permitindo dessa forma responder as duas hipóteses estudadas. Para isso, uma estimativa das toneladas de $CO_2$ emitido foi quantificado por meio da calculadora ICAO. Além dessa estimativa, três taxações foram consideradas com base nos estudos anteriores da UE, sendo assim foram considerados taxações de 10, 15 e 30 euros por tonelada de $CO_2$ emitido. Esses valores foram convertidos de euro para a cotação média do real no período, e através de uma relação estabelecida a qual considera elasticidade, preço, custo, e grau de concentração do mercado (HHI), a taxação ambiental foi transmitida para as tarifas médias das passagens aéreas (Resende e Oliveira, 2017).

Através dessa transmissão da taxação ambiental para as tarifas aéreas, foi possível analisar as duas hipóteses deste estudo. Recapitulando, a primeira hipótese é de que a reação da demanda não é constante ou linear à taxação ambiental, e de que quanto maior a taxação ambiental, maior a redução da demanda, de forma não linear. A primeira hipótese foi confirmada ao se perceber que o aumento da taxação ambiental aumentou também a queda da demanda de forma geral, porém essa queda não foi linear, inclusive devido a equação utilizada (Resende e Oliveira, 2017).

A segunda hipótese é de que o perfil dos passageiros ou da rota impactam na reação da demanda, ou seja, passageiros com diferentes motivos de viagens, e rotas com diferentes perfis de passageiros possuem impacto na demanda diferentes entre si. Para analisar essa hipótese, o mercado foi segmentado a avaliado conforme esses diferentes segmentos, o quais podem ser vistos em Resende (2016).

O segmento de maior impacto foram rotas com fator de ocupação superior a 90%, isso porque o estudo considerou perda proporcional a demanda, ou seja, quanto maior o número de passageiros em um voo, maior a perda de demanda. O fator de ocupação foi seguido pelo ano de viagem e região destino da viagem, respectivamente. O período de 2006 a 2008 foi o período de maior impacto na demanda da base de dados estudada, o que pode explicado pela crise existente no período e a ideia de que a existência de crise faz com que a demanda caia mais do que em período sem crise, conforme explicado anteriormente.

A região nordeste por sua vez, foi a região do Brasil com maior perda de demanda frente a uma taxação ambiental, e uma das explicações encontradas para isso são as características turísticas da região. Novamente, conforme abordado na explicação da elasticidade-preço da demanda, passageiros viajando a negócios são menos afetados por um aumento nas tarifas do que os passageiros a turismo, o que ajuda a explicar o motivo de queda significativa esperada na região nordeste.

## IV. CONCLUSÕES

Como visto no capítulo, emissão de gases do efeito estuda estão sido vistos cada vez mais como possível problemas, e mais pessoas estão buscando uma alternativa para redução dos mesmos, visando uma melhora ambiental do planeta (COP21; COP24; EUROSTAT, 2011). Com isso em mente, alguns países da UE começaram a avaliar uma taxação ambiental aplicada sob o consumo de combustível. Sendo assim, uma das aproximações para buscar a redução das emissões do transporte aérea seria a aplicação de uma taxação ambiental, o que motivou o estudo do ITA para voos domésticos brasileiros.

No entanto, conforme mostrado no capítulo, a implementação de uma taxação ambiental pode impactar a demanda do transporte aéreo, gerando reações não esperadas, como até mesmo o aumento das emissões no caso dos passageiros migrarem do transporte aéreo para um transporte com maiores emissões (Hofer, Dresner e Windle, 2010). Além disso, mostrou-se que a taxação ambiental impacta diretamente a demanda, fazendo com que haja uma redução da mesma. Essa redução da demanda pode ter o resultado esperado, onde as empresas aéreas buscam o equilíbrio da oferta e demanda, reduzindo frequência de voos ou se tornando mais eficientes com relação ao consumo de combustível, e consequentemente reduzindo as emissões. Porém, as empresas aéreas podem não buscar esse equilíbrio, não havendo a redução das emissões.

É importante reforçar também que a queda da demanda perante um aumento das tarifas devido a taxação ambiental ocorre de forma diferente de acordo com diferentes características do voo e do mercado. Por exemplo, regiões mais turísticas sofrem uma queda maior na demanda do que regiões mais comerciais. Vale lembrar também, que essa queda na demanda pode impactar não somente as emissões, mas também a economia regional de forma geral, podendo levar a crises financeiras em determinadas regiões do país.

Não só isso, mas a redução da demanda não é suficiente para garantir a redução nas emissões. Empresas aéreas teriam que mudar sua malha e frota para garantir o equilíbrio oferta e demanda, e não se sabe ao certo se os passageiros realmente iriam desistir da viagem reduzindo as emissões, ou se eles iriam manter a viagem, mudando apenas o meio de transporte para chegar ao destino. Neste último caso as emissões apenas são transferidas de um modo de transporte para outro, podendo haver ou não uma redução nas emissões, ou até mesmo um aumento nas emissões.

Em resumo, a pergunta inicial colocada neste capítulo foi: e se o consumo de combustível utilizado pelas aeronaves no Brasil fosse taxado por questões ambientais, quais seriam os efeitos disso no setor de transporte aéreo? Conforme visto, o principal efeito observado seria a redução da demanda, impactando diretamente as empresas aéreas. Já com relação a segunda pergunta: aplicar uma taxação ambiental sobre o consumo de combustível de voos domésticos das empresas aéreas brasileiras vai reduzir as emissões de gases do efeito estufa? A resposta a essa pergunta é que é difícil garantir que haverá essa redução das emissões. Existem diversos fatores difíceis de se controlar para garantir tal redução, tal como saber qual será a reação exata do passageiro, mudando ou não seu meio de transporte, e qual seria a reação da empresa aérea, mudando ou não sua frota e frequência de voos. Portanto, para garantir que a taxação surtiria efeito nas emissões, estudos mais aprofundados deveriam ser realizados para buscar entender a reação geral das empresas aéreas e dos passageiros.